# Zernike-efficient phase contrast with a scanned electron probe

Philipp Pelz[*]

*Institute of Micro- and Nanostructure Research (IMN) and Center for Nanoanalysis and Electron Microscopy (CENEM), Department of Materials Science and Engineering, Friedrich-Alexander-Universität Erlangen-Nürnberg (FAU), Cauerstraße 3, 91058 Erlangen, Germany*

For weak-phase objects, scanning transmission electron microscopy (STEM) with a far-field detector transfers at most half of the phase information per electron that Zernike phase-contrast (ZPC) conventional transmission electron microscopy (CTEM) achieves. We propose APEX (Apex Phase-plate EXit-wave imaging), a scanned-probe mode that combines a defocused probe, a $\frac{\pi}{2}$ phase element in a magnified image of the probe crossover, and in-focus image-plane detection at each scan position. It applies the generalized Zernike framework of Dwyer and Paganin to a scanned probe. In a linear model of a thin weak-phase object with an ideal phase dot, APEX transfers nearly the full Zernike Fisher information over a wide band, substantially more than diffraction-plane four-dimensional STEM (4D-STEM). In idealized simulations of myoglobin in vitreous ice at a cryo-EM dose (elastic scattering only, no chromatic aberration, full coherence), ideal-dot APEX at a probe defocus of 1 µm reconstructs the phase more accurately than defocused 4D-STEM ptychography with one untuned probe setting. At a defocus of about 25 µm, a crossed laser phase plate (XLPP) acts as the $\frac{\pi}{2}$ element, and APEX of a periodic specimen cell comes close to CTEM with the same plate. In two dimensions, this mode is equivalent to laser phase-plate CTEM of a small, scanned area. A single simulated APEX scan with the XLPP also separates two myoglobin phase screens 150 nm apart in vacuum, at known depths, which is not possible with one ZPC CTEM exposure.

**Practitioner points**

- With an ideal $\frac{\pi}{2}$ phase dot at the image of the probe crossover and in-focus image detection, a scanned, defocused probe (APEX) transfers nearly all of the phase information per electron that Zernike phase-contrast TEM achieves.
- A crossed laser phase plate at the crossover image acts as a $\frac{\pi}{2}$ element only if the probe is defocused so that $\alpha\Delta f \gtrsim 1400$ Å. This needs transfer optics that have not been demonstrated, and in two dimensions the mode is then equivalent to laser phase-plate TEM with a scanned illuminated area.
- One APEX scan is a limited-angle tilt series. In simulations, it separates two phase screens at known depths, which a single Zernike TEM exposure cannot do.

## I. INTRODUCTION

Because radiolytic specimen damage limits the permissible electron fluence in biological cryo-electron microscopy (cryo-EM), the fundamental figure of merit of an imaging mode is the specimen phase information extracted per incident electron[1]. Rose[2] showed already in 1974 that bright-field phase contrast in scanning transmission electron microscopy (STEM), even with optimally shaped annular detectors inside the illumination cone, requires up to about four times the dose of conventional phase-contrast transmission electron microscopy (CTEM) with parallel illumination for the same signal-to-noise ratio (SNR). For a thin weak-phase object, the quantum Fisher information (QFI) establishes the upper bound on the phase sensitivity attainable by any physical measurement. In this framework, Zernike phase-contrast (ZPC) CTEM reaches the QFI across all spatial frequencies above the phase-plate cut-on, whereas four-dimensional STEM (4D-STEM) with a detector in the diffraction plane captures only a maximum of half of the QFI for broadband illumination, with a fraction that depends on the probe, the spatial frequency and the detector configuration[3]. Bennemann *et al.* recently reached a consistent conclusion from the detective quantum efficiency (DQE), defined as the squared SNR of a reconstruction relative to that of an ideal Zernike microscope[4]: for a weak-phase object, in-focus single-sideband ptychography peaks at a DQE of only about 20%, because its signal resides in the double-overlap regions of the diffraction discs, whereas large defocus raises the low-frequency DQE through the triple-overlap regions. They also showed that 4D-STEM is considerably more robust to partial temporal coherence than ZPC CTEM, which loses its high-frequency transfer already at a defocus spread of a few nanometres unless the chromatic aberration is corrected. Varnavides *et al.* evaluated STEM phase retrieval with the spectral SNR (SSNR) of repeated reconstructions, whose square relative to that of an ideal ZPC microscope is the DQE[5]. For iterative far-field ptychography, the SSNR depends on the dose: at low fluence it approaches that of direct ptychography, and at high fluence it saturates at a value corresponding to a DQE of 1/2, consistent with the Fisher-information bound of Dwyer and Paganin[3]. For a linear reconstruction under Poisson noise, the DQE of an independently estimated weak-phase mode is bounded by its normalized linear Fisher information, which we use below (Sec. II B), and the noise-weighted matched filter reaches this bound when the mode does not couple to other modes[6] (Supplementary Note S1). Fixed-beam CTEM with a lossless phase plate is the dose-efficiency benchmark that scanned-probe methods must match[1]. It has recently been implemented[7,8] with a crossed laser phase plate (XLPP).

[*]Contact author: philipp.pelz@fau.de

APEX (Apex Phase-plate EXit-wave imaging), the scanned-probe mode we propose here, is named for the conjugate apex of the converging probe cone, where the post-specimen $\frac{\pi}{2}$ element is placed. It builds on three lines of work. First, Dwyer and Paganin recently generalized ZPC to arbitrary incident wavefields, showing that full quantum efficiency is achieved whenever a $\frac{\pi}{2}$ phase shift is applied selectively to the unscattered wave in a plane where that wave is spatially concentrated and the resulting wavefield is detected in coordinate space[9]. Their optical construction uses random phase masks before and after the specimen, a Zernike phase plate in coordinate space and a pixelated detector in the image plane, and, to avoid speckle zeros, a normalized sum over translations of the incident wave, as in STEM. They did not consider a defocused probe crossover as the plane of the phase element, an electron-optical column design, or an iterative or multi-layer reconstruction. Second, the real-space STEM configuration introduced by Etheridge and co-workers scans a focused electron probe across the specimen while recording the magnified image plane on a pixelated detector[10], but incorporates no post-specimen phase plate and performs no phase reconstruction. Third, electron ptychography provides the computational framework to stitch overlapping probe views into a single self-consistent complex object[11,12], but typically records far-field diffraction patterns rather than real-space images in the STEM.

To our knowledge, no electron-optical scheme has been reported that places the post-specimen $\frac{\pi}{2}$ element at the image of a defocused probe crossover and detects the in-focus image at each scan position. Related post-specimen phase elements have been explored in adjacent regimes; they differ from APEX in the detection plane or in the plane of the element. In hard X-ray microscopy, Vartiainen *et al.* placed a Zernike phase dot at the focal plane of a scanned beam and recorded far-field diffraction patterns for ptychographic reconstruction[13]. In fixed-beam electron microscopy, Kotani *et al.* positioned a hole-free thin-film phase plate in the selected-area aperture plane conjugate to the condenser crossover[14,15]. In confocal electron microscopy, Nagayama considered inserting spatial filters or phase plates into both the illumination lens and the outgoing projection lens (Claim 1 of Nagayama[16]). In visible-light microscopy, two patents describe related arrangements. Minobe *et al.* placed a phase plate at the convergence point behind a specimen under a fixed converging beam and viewed the full-field image, without a scan[17]. Kimura described a confocal scanning phase-contrast microscope with a detection-side phase plate; its light spot is focused on the specimen, and a point detector behind a pinhole records one signal per scan position[18]. Related pre-specimen approaches include scanning X-ray Zernike microscopy by Helmholtz reciprocity, which places a phase ring before the probe-forming lens and records one integrated intensity per scan position[19], and matched illumination and detector interferometry (MIDI-STEM), which introduces a phase plate into the probe-forming aperture[20].

In our proposed geometry (Fig. 1), a defocused electron probe illuminates a local patch of the specimen, the objective lens images the probe crossover onto a $\frac{\pi}{2}$ phase-shifting element located in a magnified conjugate plane of the crossover, and a pixelated detector records the magnified in-focus image of the illuminated patch at each scan position. Because the analysis below introduces several new concepts, we summarize our contributions in this work upfront:

1. We identify the defocused probe crossover as a plane in which the unscattered wave is tightly concentrated while the scattered waves are spatially separated. A finite phase element in this plane approximates the Zernike projector of Dwyer and Paganin[9] in a scanned-probe geometry. We show that the separation does not depend on the magnification of the plane in which the phase element resides (Sec. II C).
2. We evaluate the quantum and the linear Fisher information per electron for scanned geometries (Sec. II A, Sec. II B).
3. We show that, for the same $\frac{\pi}{2}$ dot at the crossover image, conjugate image-plane detection transfers nearly twice the Fisher information per electron of far-field diffraction detection (Sec. IV A).
4. We compare laser phase plates and an idealized electrostatic point charge on the focused spot, whose projected potential imparts a logarithmic phase, with an ideal $\frac{\pi}{2}$ dot, and quantify the effect of image-side spherical aberration for the ideal dot (Sec. IV B, Sec. IV C).
5. We formulate the instrumental requirements of APEX, showing that a standard microscope column supports APEX only at small convergence angles, whereas APEX at milliradian convergence angles requires an additional magnifying stage such as a transfer-lens column (Sec. IV D).
6. We benchmark APEX on simulated myoglobin in vitreous ice at a cryo-EM dose. At a probe defocus of 1 μm with an

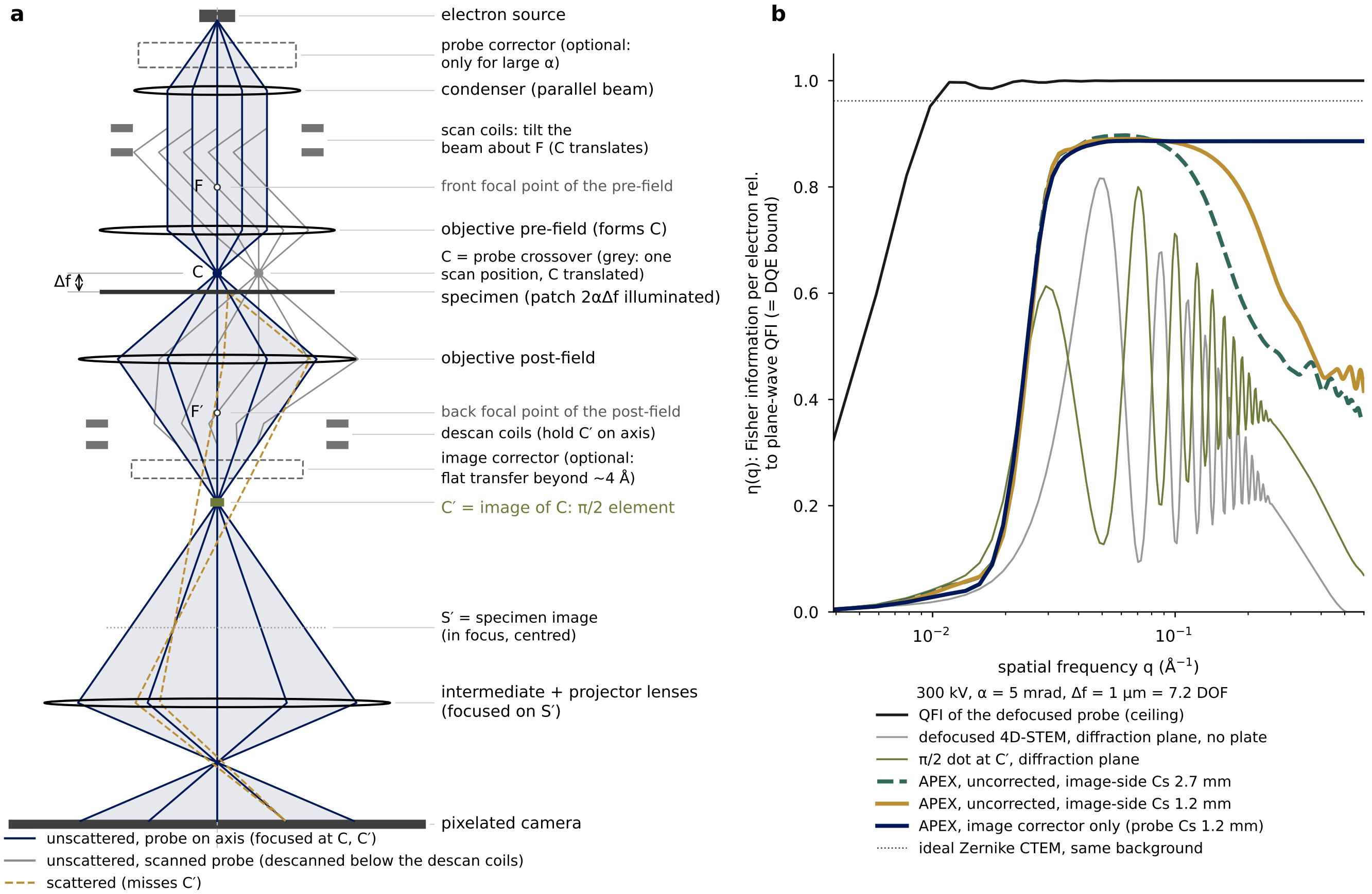

FIG. 1. Optical geometry and Fisher information transfer of APEX. (a) Paraxial ray trace of the column (not to scale). The objective pre-field focuses the beam to the probe crossover $C$, and the post-field images $C$ to $C'$, where the $\frac{\pi}{2}$ element shifts the focused unscattered wave. The scan coils translate $C$ (grey: one scan position), and the descan coils keep $C'$ on the element. (b) Linear Fisher information per electron $\eta(q)$ (Eq. (2)), relative to the plane-wave QFI. Black: QFI of the defocused probe (Eq. (1)). Dotted: ideal ZPC CTEM. Supplementary Note S2 gives the parameters.

ideal dot, it reconstructs the phase more accurately than defocused 4D-STEM ptychography. At a probe defocus of about 25 μm with an XLPP, it comes close to plane-wave CTEM with the same plate. Defocused CTEM without a phase plate is included as a baseline (Sec. IV E).

7. We implement a differentiable image-plane forward model (Sec. V) and formulate a joint linear Wiener solver that inverts multiple specimen layers simultaneously and applies identically to APEX and 4D-STEM data (Sec. VI A).
8. We show that, because each APEX frame is a point projection from the crossover, a single scan records a limited-angle tilt series with ZPC, allowing us to separate two myoglobin phase screens spaced 150 nm apart at known depths, which a single ZPC CTEM exposure cannot separate (Sec. VI, Sec. VI B).

## II. THEORY

### A. Quantum Fisher information ceiling

The QFI has two uses in this work. First, the QFI of a plane wave is the unit of all efficiencies $\eta$ (Eq. (2)): a value of 1 means that a detection scheme extracts all the phase information that one electron carries. Second, the QFI of the defocused probe is the upper limit for any detector behind this probe, and we compare APEX and 4D-STEM with this limit (black curve in Fig. 1(b)). No reconstruction uses the QFI.

Let $\psi_0(\boldsymbol{r})$ denote the incident electron wavefunction at the specimen plane, normalized to a single incident electron ($\int |\psi_0(\boldsymbol{r})|^2 \, \mathrm{d}\boldsymbol{r} = 1$), and consider a thin phase object with transmission function $\exp(i\sigma\varphi(\boldsymbol{r}))$, where $\sigma$ is the interaction constant and $\varphi$ the projected potential, yielding the exit wave $\psi(\boldsymbol{r}) = \psi_0(\boldsymbol{r}) \exp(i\sigma\varphi(\boldsymbol{r}))$. Because the derivative of the exit wave with respect to any real parameter $\theta$ of the projected potential $\varphi$ is $\partial_\theta \psi = i\sigma(\partial_\theta \varphi)\psi$, the QFI of a pure state, $4\big(\langle \partial_\theta \psi \mid \partial_\theta \psi \rangle - |\langle \psi \mid \partial_\theta \psi \rangle|^2\big)$ (Eq. 26 of Paris[21]), is per electron the intensity-weighted spatial variance of the parameter derivative across the probe footprint:

$$J_\theta = 4\sigma^2 \mathrm{Var}_{|\psi_0|^2}(\partial_\theta \varphi). \quad (1)$$

For a Fourier mode of $\varphi$ at spatial frequency $\boldsymbol{q}$, averaged over its cosine and sine quadratures, Eq. (1) evaluates to $4\sigma^2\left[\frac{1}{2} - \frac{|\tilde{I}(\boldsymbol{q})|^2}{2}\right]$ per mode, where $\tilde{I}(\boldsymbol{q})$ is the Fourier transform of the incident probe intensity $|\psi_0(\boldsymbol{r})|^2$ (Eq. 16 of Dwyer and Paganin[3] in our notation). The QFI depends only on the illumination intensity at the specimen: a uniform plane wave maximizes the QFI across all non-zero spatial frequencies, whereas a finite probe experiences an information roll-off only at spatial frequencies $q$ below the inverse diameter of the illuminated patch. Above this spatial frequency, a defocused probe thus loses almost no information by its illumination. The loss of a scanned-probe mode comes from the detection, which Sec. II B quantifies.

### B. Linear Fisher information relative to a known reference wave

To compare practical detection geometries on an equal footing, we evaluate the classical Fisher information that is linear in the weak phase $\varphi$ (homodyne detection against the known unscattered wave) averaged over cosine and sine modes at spatial frequency $q$. It is the classical Fisher information of a pure state under a projective measurement (Eq. 5 of Dwyer and Paganin[3]; Koppell and Kasevich[22] give the general Poisson form), to which we add the background $B$ and the normalization:

$$\eta(q) = \frac{1}{2\cdot 2N_p} \sum_{p,\ \mathrm{pixels}} \frac{4(\mathrm{Re}[r^* d])^2}{|r|^2 + B}. \quad (2)$$

Here, $r = U\psi_0$ denotes the unscattered reference wave at the detector plane, $d = U\big(i f_{\boldsymbol{q}}\psi_0\big)$ is the first-order scattered wave for cosine and sine modes $f_{\boldsymbol{q}}$ at spatial frequency $q$, $U$ is the unitary or non-unitary transfer operator of the post-specimen optics, $N_p$ is the number of probe positions, and $B$ is a background intensity that accounts for electrons scattered by the surrounding vitreous ice. This background model is used for all Fisher-information calculations (Sec. IV A, Sec. IV B, Sec. IV C); the reconstructions in Sec. IV E and Sec. VI B instead contain the ice explicitly in the simulated specimen. A background of 4% corresponds to thin ice. With the apparent mean free path of about 320 nm for scattering outside the objective aperture[23], it corresponds to about 13 nm of ice. The weak-phase model of our simulated specimen scatters less and gives about 30 nm. Thicker ice gives a larger background. Between background fractions of 0 and 20%, the band values of 4D-STEM fall by at most 0.06. Those of APEX and of ZPC CTEM fall approximately as $1/(1+s)$ for a background fraction $s$ (Table S7).

Eq. (2) is the Poisson Fisher information $\sum (\partial_\theta \mu)^2/\mu$ of the expected intensity $\mu = |r + \theta d|^2 + B$, kept to first order in $\theta$, so that $\partial_\theta \mu = 2\ \mathrm{Re}[r^* d]$. In the prefactor, one factor of 2 averages the cosine and sine modes, the other divides by the per-mode ZPC value 2 from Eq. (1) (with $\sigma = 1$), and $N_p$ averages over the probe positions. All efficiencies $\eta$ in this work are therefore given relative to the background-free ZPC value, i.e. the QFI of a plane wave; with the 4% background, ideal plane-wave ZPC CTEM reaches $\eta = 0.962$ above its cut-on. Eq. (2) is the Fisher information at zero object phase. It omits the dark-field term $|d|^2$, whose derivative vanishes at $\theta = 0$ but which carries information for a finite object phase[3]; we assess the finite-object performance by the reconstructions in Sec. IV E and Sec. VI B.

### C. Optimal axial placement of the $\frac{\pi}{2}$ phase element

Dwyer and Paganin[9] showed that generalized ZPC imaging with an arbitrary incident state $\psi_0$ is obtained with the post-specimen operator $\hat{Z} = 1 + (i-1)\ |\psi_0\rangle\langle\psi_0|$ followed by detection in coordinate space (their Eq. 26), and identified the physical implementation of the rank-one projector $|\psi_0\rangle\langle\psi_0|$ as the main difficulty. We approximate this projector by a finite $\frac{\pi}{2}$ element in a plane where the unscattered wave is tightly focused while the scattered waves are spatially separated; the simulations below quantify how closely this approximation approaches the ideal operator. For a probe defocused by an axial distance $\Delta f$ from the specimen, the unscattered wave focuses to an Airy spot at the probe crossover, whereas a wave scattered by specimen spatial frequency $q$ forms a laterally shifted Airy spot displaced by $\lambda\Delta f q$ in the crossover plane (Fig. 2 a). Expressed in units of the Airy disk radius $0.61\frac{\lambda}{\alpha}$, the dimensionless separation between the scattered and unscattered spots at the crossover is

$$s = \frac{\Delta f\, q\, \alpha}{0.61}, \quad (3)$$

which is invariant to the optical magnification of the conjugate plane $C'$ in which the physical phase element resides. Choosing a phase dot of radius equal to two Airy units places the cut-on spatial frequency near $q_c \approx \frac{1.22}{\alpha\Delta f}$, which corresponds to 2.44 divided by the illuminated patch diameter $2\alpha\Delta f$ at the specimen (Fig. 2 b). Conversely, if the probe is focused directly onto the specimen ($\Delta f = 0$), the crossover coincides with the specimen itself and the unscattered wave fills the bright-field disk in the back focal plane, shifting the cut-on frequency to $2\frac{\alpha}{\lambda}$.

## III. METHODS

To evaluate Eq. (1) and Eq. (2) numerically across realistic optical configurations, we implemented a Fourier-optics Fisher information calculator on a periodic grid (Table S1 lists the parameters of all simulations). For each spatial frequency mode, the exit wave is propagated through the complete post-specimen optical chain: objective-lens aberrations and a back-focal-plane phase plate (when present), both applied to the spectrum of the exit wave, Fresnel propagation by $-\Delta f$ to the conjugate crossover image $C'$, multiplication by the complex transmission of the $\frac{\pi}{2}$ phase element, Fresnel propagation by $+\Delta f$ back to the conjugate specimen image plane $S'$, and, where stated, a detector defocus. Applying the aberrations before the crossover image corresponds to an objective lens that forms the magnified intermediate images, including $C'$, behind its aberrated back focal plane. Here and in the Fisher-information calculations, $\Delta f > 0$ places the crossover upstream of the specimen, as in Fig. 1; the thin-specimen simulations of Sec. IV E place it downstream, which changes the band-averaged transfer of the ideal dot by at most 0.02. We verified that this calculation reproduces the analytical $\sin^2$ contrast transfer function of defocused CTEM, the ideal ZPC CTEM limit, and the analytical disk-overlap expressions for in-focus 4D-STEM to within the discretization error. The electrostatic element is an ideal projected point charge centred on the unscattered spot, which imparts the logarithmic phase $\varphi(r) = -A \ln \max(r, r_c)$ with a flat core of radius $r_c$ and no support. Continuous-wave laser phase plates are modelled as standing waves in a plane conjugate to the back focal plane, each a Gaussian phase stripe with the numerical aperture (NA) and effective focal length of Petrov *et al.*[7] (Table S1). As in both instruments[7,8], the polarization is horizontal, so we model the $\cos^2$ standing-wave fringes explicitly. A fringe-free plate at the relativistic reversal angle retains only the mean phase, which is half of the antinode phase[24], and therefore requires twice the power for the same phase shift. The single laser phase plate (SLPP) has $\frac{\pi}{2}$ at its central antinode, as demonstrated by Petrov *et al.*[25]. The XLPP proposed by Petrov *et al.*[7], which Yu *et al.*[8] have since used for cryo-EM imaging, crosses two orthogonal, non-interfering standing waves. Each imparts $\frac{\pi}{4}$ at its antinodes, so that the unscattered beam receives $\frac{\pi}{2}$ where the two central antinodes cross. By Eq. 3 of Petrov *et al.*[7], this requires a circulating power of 38 kW per cavity, comparable to the powers at which Yu *et al.*[8] measured a total phase shift close to $\frac{\pi}{2}$. In APEX, the plate is placed at $C'$ with $M_C \Delta f = f$, which gives the same scale $r_{C'} = f \lambda q$ between spatial frequency and position on the plate as in CTEM; the finite crossover spot, however, averages the plate profile. The axial offset between the two standing waves[8] is negligible at $C'$, because the unscattered wave converges there with the small semi-angle $\alpha / M_C$ (Table S1).

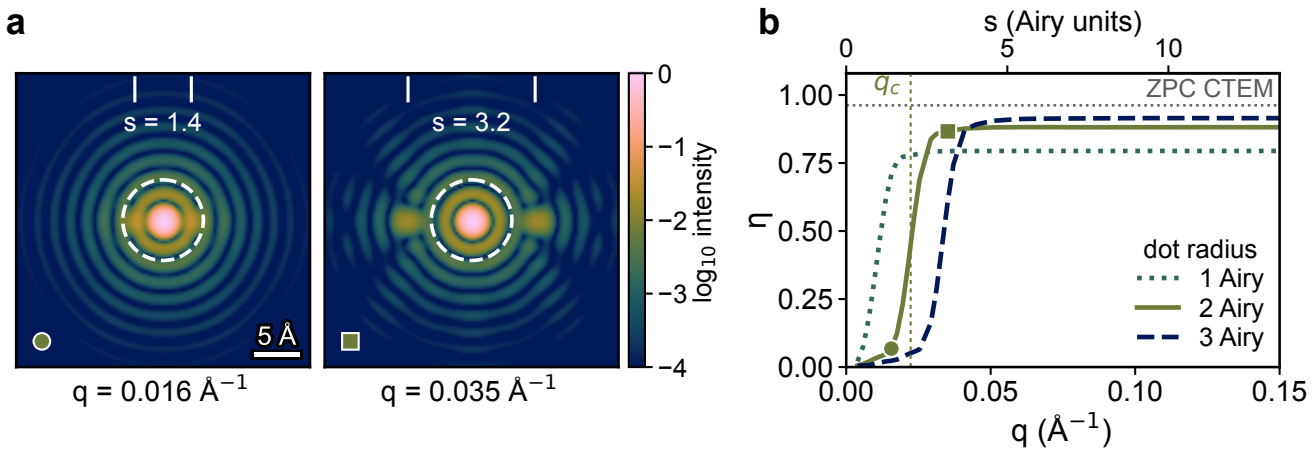

FIG. 2. Placement of the $\frac{\pi}{2}$ element at the probe crossover. (a) Intensity at the crossover after a cosine phase grating at $q = 0.016$ Å⁻¹ (left, circle) and $q = 0.035$ Å⁻¹ (right, square). The first-order spots (white ticks) fall inside the dot (dashed) at $s = 1.4$ and pass outside it at $s = 3.2$. (b) Linear Fisher information per electron $\eta$ (Eq. (2)) of APEX for dot radii of 1, 2 and 3 Airy units, against $q$ and against the separation $s$ of Eq. (3) (top axis). Markers: the two frequencies of (a). Dotted: ideal ZPC CTEM. Supplementary Note S2 gives the parameters.

## IV. RESULTS

### A. Dependence of Fisher information transfer on the detection plane

To isolate the role of the detector plane, we first compare diffraction-plane and image-plane detection with and without a post-specimen $\frac{\pi}{2}$ phase dot (Fig. 1 (b), Fig. S2). In the absence of a phase plate, conventional in-focus 4D-STEM with a diffraction-plane detector transfers less than 0.2 of the ideal ZPC information per electron at all spatial frequencies (Fig. S2 (a)), because electrons scattered outside the bright-field disk do not interfere with the unscattered reference beam. This value for a single in-focus probe lies below the estimate of about one half for broad-band illumination[3] and is consistent with the peak DQE of about 20% that Bennemann *et al.* report for in-focus single-sideband ptychography[4]. Moving the detector to a defocused image plane keeps the scattered and unscattered waves spatially overlapped and recovers the oscillatory contrast transfer of defocused CTEM, with zeros of the contrast transfer function (Fig. S2 (a)). With an ideal $\frac{\pi}{2}$ phase dot of radius 2 Airy units at the crossover image $C'$ of a defocused probe, far-field diffraction detection, the X-ray geometry of Vartiainen *et al.*[13], transfers less than half of the ZPC bound and oscillates with $q$. Detection in the in-focus conjugate image plane $S'$ (APEX) removes the oscillations and transfers 0.88 above the cut-on of the dot (Fig. 1 (b)). The detector must stay in the image plane: 50 nm of detector defocus behind the dot brings back the zeros above 0.1 Å⁻¹ (Fig. S2 (a)). Conversely, if the probe is focused directly onto the specimen ($\Delta f = 0$) and a $\frac{\pi}{2}$ plate covers the entire bright-field disk, image-plane detection reaches the ZPC value only above the cut-on $2\alpha/\lambda$, which moves to higher $q$ as $\alpha$ increases (Fig. S2 (b)).

### B. Comparison of practical $\frac{\pi}{2}$ phase elements at the crossover image

To assess how closely physically realizable phase plates approach an ideal circular $\frac{\pi}{2}$ dot at $C'$, we compared single and crossed continuous-wave laser phase plates with an electrostatic point charge on the unscattered spot (Fig. 3, Fig. 4). A laser standing-wave phase plate puts no solid material into the electron beam path, so the plate itself adds no inelastic scattering and no contamination at the crossover image. With the SLPP at $\Delta f = 1$ μm, the Gaussian laser stripe shifts both the focused unscattered spot and those scattered waves propagating parallel to the laser axis, so that almost no information is transferred along the laser axis. Its fringes are much finer than the spot and average the phase on the spot to $\frac{\pi}{4}$, so that the transfer stays below half of the ZPC value even perpendicular to the stripe (Fig. 3 a). The XLPP[7,8] removes this blind direction: each of its two orthogonal stripes imparts $\frac{\pi}{4}$, so that the unscattered spot at the crossing receives $\frac{\pi}{2}$ and a wave scattered along either laser axis still acquires a phase difference of $\frac{\pi}{4}$ (Fig. 4 b). Each of its two non-interfering standing waves retains full-contrast $\cos^2$ fringes along its axis, and where the two stripes cross, the fringes form a square lattice of antinodes with the period $(\lambda_l/2)\Delta f/f$ at $C'$ in specimen units[7] (Fig. 4 a). At $\Delta f =$ 1 µm, this period is much smaller than the unscattered spot, which therefore receives only the fringe-averaged phase, about $\frac{\pi}{4}$ at the power per cavity reached by Yu *et al.*[8]. The higher Kapitza–Dirac orders of the lattice form ghost images far from the patch that carry almost no information, and the transfer stays below half of the ZPC value (Fig. 3 a). Twice that power, close to the power of the SLPP[25], restores $\frac{\pi}{2}$ on average and raises the transfer, but not to that of the ideal dot (Fig. 3 a). Because the phase scales with the product of numerical aperture (NA) and power (Eq. 3 of Petrov *et al.*[7]), the larger NA anticipated for the XLPP[7] would lower this requirement to about 50 kW per cavity. At $\Delta f = 25.4$ µm, the lattice period grows larger than the spot, which sits inside a single antinode crossing that imparts the full $\frac{\pi}{2}$. All laser plates then come close to the ideal dot at the same defocus (Fig. 3 b); only the SLPP keeps its weak direction along the laser axis. This motivates the large defocus used with the XLPP in Sec. IV E. The transfer lenses set the effective focal length $f = M_C\Delta f$ at $C'$, but $M_C$ scales the spot and the offsets $\lambda\Delta f q$ of the scattered waves equally. When $f$ is reduced until the spot fits a single antinode, the stripes cover all spatial frequencies below $q_w \approx 1.55/(\mathrm{NA}\ \alpha\Delta f)$. The laser plates therefore require $\alpha\Delta f \gtrsim 1.55/(\mathrm{NA}\ q_{\min})$, about 1400 Å for $q_{\min} = 0.02$ Å$^{-1}$, independently of $\alpha$: at this $\alpha\Delta f$, the XLPP transfers the same at 10 and 15 mrad (Fig. 3 b), whereas at $\Delta f = 1$ µm no choice of $f$ brings the XLPP close to the ideal dot. At $\Delta f = 1$ µm, a point charge centred on the spot transfers more than either laser plate in every band. Its projected potential imparts the logarithmic phase $-A\ln r$, which is azimuthally symmetric, so it has no blind direction: a scattered wave of spatial frequency $q$ passes $C'$ at $r = \lambda\Delta f q$ and acquires the phase difference $A\ln(\lambda\Delta f q/r_c)$ (Fig. 4 b) of the unscattered spot relative to the scattered wave, which grows only logarithmically with $q$. For a core radius of half an Airy radius, the transfer approaches that of the ideal dot at intermediate spatial frequencies and falls off on both sides, because the phase difference passes through $\pi/2$ only once (Fig. 3 a). A smaller core changes the transfer little, and the transfer is nearly independent of the direction of $\boldsymbol{q}$.

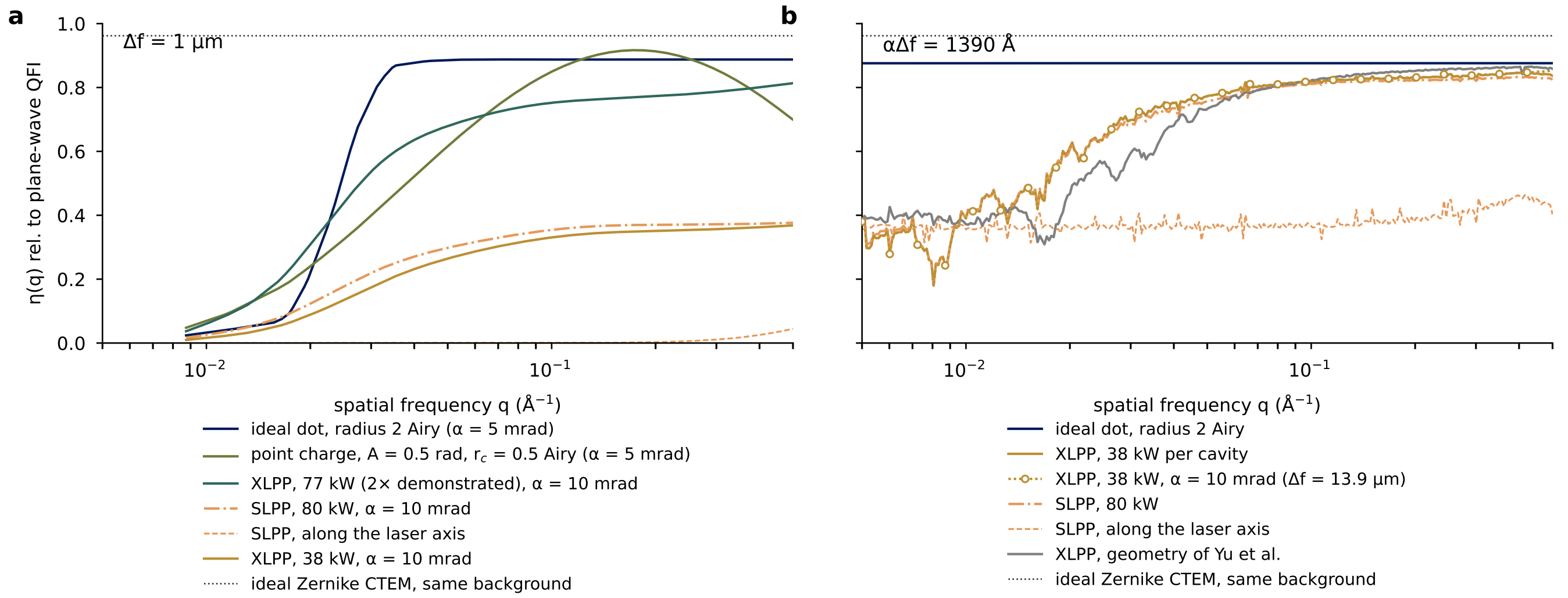

FIG. 3. Linear Fisher information per electron $\eta(q)$ of the $\frac{\pi}{2}$ elements at the crossover image $C'$, relative to the plane-wave QFI, averaged over the directions of $\boldsymbol{q}$. (a) $\Delta f = 1$ µm. (b) $\alpha\Delta f = 1390$ Å. Dashed: the SLPP along its laser axis, its worst direction. Dotted: ideal ZPC CTEM. Table S1 and Supplementary Note S2 give the parameters.

### C. Electron-optical aberration requirements

At the moderate convergence semi-angles ($\alpha \approx 5$ mrad) suited to a physical $\frac{\pi}{2}$ phase element, APEX does not need a spherical-aberration ($C_s$) corrector on the probe-forming side of the objective lens (Fig. 1 (b)). With an image-side corrector alone, an uncorrected probe changes the transfer of the double-corrected column by at most 0.01 in each band. On the image-forming side, uncorrected spherical aberration, balanced by a Scherzer-like objective defocus, leaves the transfer below 0.1 Å$^{-1}$ unchanged and lowers it at higher spatial frequencies; the loss is larger for the $C_s$ of the transfer lenses of Yu *et al.*[8] than for a standard objective (Fig. 1 (b)).

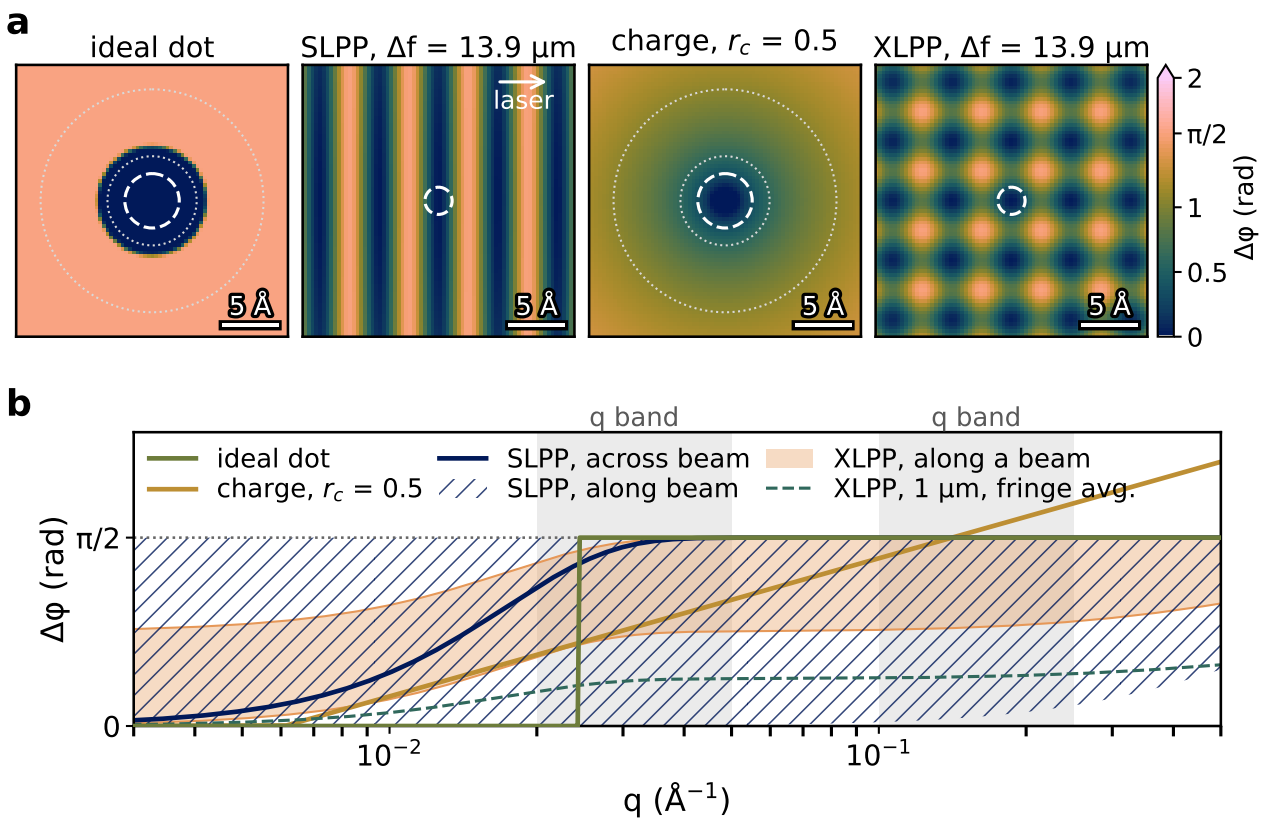

FIG. 4. The $\frac{\pi}{2}$ elements of Fig. 3 at the crossover image $C'$, referred to the specimen. (a) Phase $\Delta\varphi(\boldsymbol{r})$ of the unscattered spot relative to a scattered wave that passes the plane at $\boldsymbol{r}$. Dashed: Airy radius of the probe; dotted: $r = \lambda\Delta f q$ for $q = 0.02$ and 0.05 Å$^{-1}$. (b) $\Delta\varphi$ against $q = r/(\lambda\Delta f)$. Hatched and orange: range spanned by the standing-wave fringes along the SLPP beam and along one XLPP beam. Dashed: the XLPP at $\Delta f = 1$ µm. Grey: the bands 0.02–0.05 and 0.1–0.25 Å$^{-1}$. Supplementary Note S2 gives the parameters.

This does not hold at $\alpha = 15$ mrad, the convergence semi-angle of the large-defocus depth benchmark (Sec. VI B). There, an uncorrected objective ($C_s = 1.2$ mm) on either side lowers the transfer in every band (Table S6). With a residual $C_s$ of up to 25 µm on both sides, the transfer stays within 0.03 of that of the corrected column. The 15 mrad benchmark therefore assumes spherical-aberration correction on both sides of the objective lens. The translated-crossover scan of Fig. 1 moves the crossover image across the plane of $C'$ by the scan distance times $M_C$, so that post-specimen descan coils are necessary to keep the crossover on the phase plate. In the Fisher-information model, a static displacement of the dot from the focused spot by up to 0.5 Airy units changes the band values by less than 0.02. A displacement of one Airy unit lowers the lowest band by about 0.1. At two Airy units, where the spot lies on the edge of the dot, the transfer collapses (Table S6). Referred to the specimen, 0.5 Airy units are 0.12 nm at $\alpha = 5$ mrad; in the plane of $C'$, the tolerance is larger by the factor $M_C$. Pivoting the scan at $C$ would avoid the descan, but it tilts the beam in the objective lens by an amount that depends on the scan position, which introduces coma and astigmatism on an uncorrected column, and it removes the parallax needed for depth sectioning (Sec. VI).

### D. Instrumental requirements

The key instrumental requirement of APEX is a highly magnified image of the probe crossover; Table S2 collects the estimates of this section. Because the $\frac{\pi}{2}$ element sits in a plane conjugate to the crossover $C$, the magnification $M_C$ of this plane must be large enough for the element to resolve the focused unscattered spot, whose specimen-referred Airy radius $0.61\lambda/\alpha$ becomes $M_C \cdot 0.61\lambda/\alpha$ at the element. For the point charge of Fig. 3, $M_C$ does not change the logarithmic phase profile, since $\ln r$ is scale-free: a charge $Q_{\rm el}$ at $C'$ produces $A = \sigma Q_{\rm el}/(2\pi\varepsilon_0)$ at any magnification, so that a few tens of elementary charges suffice. The charge must, however, remain within about 0.5 Airy radii of the spot, which requires $M_C \geq 10^3$ for a typical source size. For a laser element, the laser optics fix the stripe width and the fringe period, so that the scale $M_C\Delta f$ of the crossover plane must match the effective focal length of the laser geometry. The unscattered spot then fits inside a single antinode, and the stripes leave all $q \geq 0.02$ Å$^{-1}$ free, when $\alpha\Delta f \gtrsim 1400$ Å (Sec. IV B).

In a standard column, refocusing the objective lens by $\Delta f$ places $C'$ in the selected-area aperture plane, where $M_C$ equals the objective magnification at that plane, about 50× on a Titan Krios[26]. This is too small to resolve the spot at $\alpha = 5$ mrad, so that a standard column supports APEX only at $\alpha \lessapprox 0.3$ mrad (Table S2). Because the product $\alpha\Delta f$ remains unchanged, the two-dimensional transfer is unaffected: at a fixed $\alpha\Delta f$, the point charge transfers the same within about 0.01 at $\alpha$ = 0.1, 0.3 and 5 mrad. We did not simulate the myoglobin reconstruction at these small angles. The small $\alpha$, however, widens the geometric axial half-width $0.35/(\alpha q)$ (Sec. VI) by more than an order of magnitude. APEX with single-scan depth sectioning therefore requires an additional magnifying stage with an accessible plane, such as the transfer-lens column built for the laser phase plate[25]. In either column, the phase of the element must remain stable during the scan, and the descan must hold $C'$ on the element to within a fraction of the Airy radius.

The camera must record the in-focus image of the illuminated patch at each scan position. Because the linear signal arises from the interference of the reference wave (angles up to $\alpha$) with the scattered waves (up to the collection semi-angle $\theta_{\rm ap}$), its spatial frequencies extend to $(\alpha + \theta_{\rm ap})/\lambda$, and Nyquist sampling requires a specimen-referred pixel size $p \approx \lambda/[2(\alpha + \theta_{\rm ap})]$, independent of the defocus. The frame must cover the patch of diameter $2\alpha\Delta f$, together with a margin for the waves diffracted by the phase element, which we estimate as about $0.8\alpha\Delta f$ on each side from the diffraction angle of a dot of radius 2 Airy units. At $\Delta f = 1$ µm, a frame of a few hundred pixels across is sufficient. At $\Delta f = 25.4$ µm, as required by the fringed laser plate, the patch alone fills a standard 4,096-pixel cryo-EM camera, but the margin does not fit (Table S2); we have not tested how much a cropped margin costs. The number of pixels per frame thus scales with $(\alpha\Delta f)^2$, whereas the number of scan positions needed to cover a field scales inversely with the patch area, so that, at fixed overlap and relative margin, the total number of recorded pixels per specimen area is approximately independent of the defocus. Large-defocus APEX therefore records a few large frames, as a conventional cryo-EM camera does, whereas small-defocus APEX with an SLPP, a point charge or an ideal dot records many small frames and requires a fast pixelated detector, as used for 4D-STEM. As an example, we estimate the acquisition of 1 µm² at 30 e/Å² (Table S2). At $\Delta f = 1$ µm, we take a scan step of 0.56 patch radii. The field then needs about $10^5$ scan positions, each with a frame of about 300 pixels. A detector with the Timepix4 chip of Llopart *et al.*[27] has 448 × 512 pixels, which is enough for this frame. In its event-driven mode, such a detector records each electron with its arrival time, so that no frame rate limits the scan. We assume a largest beam current of 50 pA. The exposure then takes about 10 s, and this time does not depend on the scan step. The dwell time is about 90 µs per scan position. With a third of the scan step, the field needs about $10^6$ positions and the dwell time is about 10 µs. Both dwell times are far above the shortest dwell time, for which we assume 200 ns. Ding *et al.*[28] measured the detection efficiency of this chip at 100 and 200 kV. Its efficiency at 300 kV is not known to us. At $\Delta f = 25.4$ µm, the frame is larger than this detector, and a large counting camera is necessary. With a step of 0.56 patch radii, about 160 frames cover the field. The frame rate is not the limit there. At a dose rate of 12 electrons per pixel and second, which is between the two rates that Sun *et al.* tested for a counting camera[29], the exposure takes about 21 s. Conventional CTEM with the same camera and pixel size needs about 15 s for the same field. These estimates do not include the settling time of the scan and descan coils or the data transfer.

Two laser-phase-plate microscopes now exist on which parts of APEX could be tested. In the instrument of Petrov *et al.*[25], a spherical-aberration corrector follows the objective lens, and transfer optics relay a magnified image of the back focal plane onto a single-cavity laser phase plate with $f_{\mathrm{eff}} = 14.1$ mm. The XLPP of Yu *et al.*[8] sits behind transfer lenses with $f = 6.8$ mm. Several operations that APEX requires are already performed on these instruments during laser alignment. To align the laser, the illumination is converged with the condenser, the standing wave is imaged in a Ronchigram, and a pair of deflectors above and below the phase plate translates the focused unscattered beam onto the antinode[25]; this is the operation of placing a crossover image on the element. The same deflector pair is the natural candidate for the descan of Fig. 1. Petrov *et al.* report, however, that deflecting the beam by a few micrometres in the diffraction plane already increases the aberrations through the corrector. Beam-image shift acquisition has not yet been implemented with either the SLPP or the XLPP[8,25]. The shorter focal length of the plate of Yu *et al.* halves the $\alpha\Delta f$ that the antinode condition requires (Table S2). At this $\alpha\Delta f$, however, the stripes cover the spatial frequencies below 0.05 Å⁻¹ (Sec. IV B), and $q \geq 0.02$ Å⁻¹ requires $\alpha\Delta f \gtrsim 1400$ Å with either focal length. A two-dimensional APEX scan at $\Delta f \gtrsim 25$ µm could therefore be realized with the present optics: it needs a small $\alpha$, a crossover far above the specimen and its image close to the back-focal-plane conjugate. This mode is, however, equivalent to laser phase-plate CTEM with a small, scanned illuminated area, so it gives no two-dimensional benefit over that mode. What APEX adds to that mode is single-scan depth sectioning at $\alpha$ of a few milliradians, which we show only for two prescribed phase screens (Sec. VI B). It places the crossover only 10–25 µm above the specimen, so that its image lies close to the first image plane rather than the back focal plane. Realizing it requires transfer optics that image this plane onto the laser with $M_C \approx f/\Delta f \approx 600$, together with a descan that holds $C'$ on the antinode during the scan without displacing the beam in the corrector. Whether the existing transfer lenses can be excited in this way is not documented.

### E. Simulated reconstruction of myoglobin in vitreous ice at 30 e/Å²

To test whether the theoretical Fisher information advantage of APEX translates into improved two-dimensional phase micrographs of biological macromolecules, we simulated the atomistic sperm-whale myoglobin specimen from our companion cryo-ptychography study[30]: myoglobin molecules embedded in molecular-dynamics-relaxed amorphous ice, imaged with an absorptive potential by multislice, assuming an energy filter that retains only elastically scattered electrons (Table S1). None of these simulations includes chromatic aberration, objective spherical aberration, partial spatial coherence, detector modulation transfer, or the aberrations of the optics that magnify $C'$. The focused spot is also exactly centred on the element at every scan position: we did not simulate descan errors or drift of the element (Sec. IV C gives a Fisher-information estimate for a static displacement). Fig. 6 reports one Poisson shot-noise realization; for the principal modes, we repeated the simulation and the maximum-likelihood (ML) reconstruction with two further independent realizations (Table S3). The spread over these realizations shows shot noise only: all runs use the same ice, the same particles, one dose and one scan geometry. All compared imaging modes share the same solvated specimen, total electron dose, and maximum collection semi-angle. Our 4D-STEM baseline is the defocused iterative ptychographic reconstruction of the same micrograph. It uses one probe setting, which is not the best setting for 4D-STEM. We scanned the Fisher information of 4D-STEM over 39 probe settings (Fig. S3 (a), (b)). The mean of the four band values rises from 0.32 at our setting to 0.43 at the best setting. No setting reaches the band value of APEX in any band (Table S6). We also reconstructed one thin layer with one linear Wiener solver for 4D-STEM, APEX and ZPC CTEM (Table S8). In this test, 4D-STEM improves with the defocus up to 3 µm, the largest value that the simulated field allows. At 3 µm, its correlations after the 20 Å and the 10 Å low-pass reach those of APEX at 1 µm. APEX at the same defocus stays above 4D-STEM in every column of Table S8. Above 0.05 Å$^{-1}$, 4D-STEM stays below APEX at every tested setting. At low spatial frequencies, a 4D-STEM experiment with a larger defocus can therefore exceed the 4D-STEM values of Fig. 6. APEX uses the same incident probe and scan grid, a $\frac{\pi}{2}$ phase element at the crossover image $C'$ (either a circular dot of radius 2 Airy units or the XLPP described below), a conjugate image-plane camera, and the same optimization settings (Poisson ML loss, pure-phase object constraint, Adam). In these simulations, the dot has a cosine-squared transition in complex transmission, which attenuates the wave within the edge; the Fisher-information and depth calculations use a sharp, lossless dot. For comparison with fixed-beam imaging, we also simulated plane-wave CTEM exposures at the same dose and reconstructed their phase both via the same iterative Poisson ML solver and via a one-step linear Wiener filter. The scanned and plane-wave ML optimizations differ in two ways. First, the scanned reconstructions update the object after each minibatch of probe positions, whereas the full-frame CTEM optimization performs one gradient step per epoch. Second, the CTEM phase estimate is explicitly band-limited to the objective aperture, whereas the scanned object arrays are regularized implicitly through the band-limited forward model. A part of the score differences between the modes can therefore come from the reconstruction settings: we compared the modes with one linear solver and one prior only in the depth benchmark (Sec. VI B). All reconstructed micrographs are evaluated directly against the ground-truth projected phase (Fig. 5, Fig. 6).

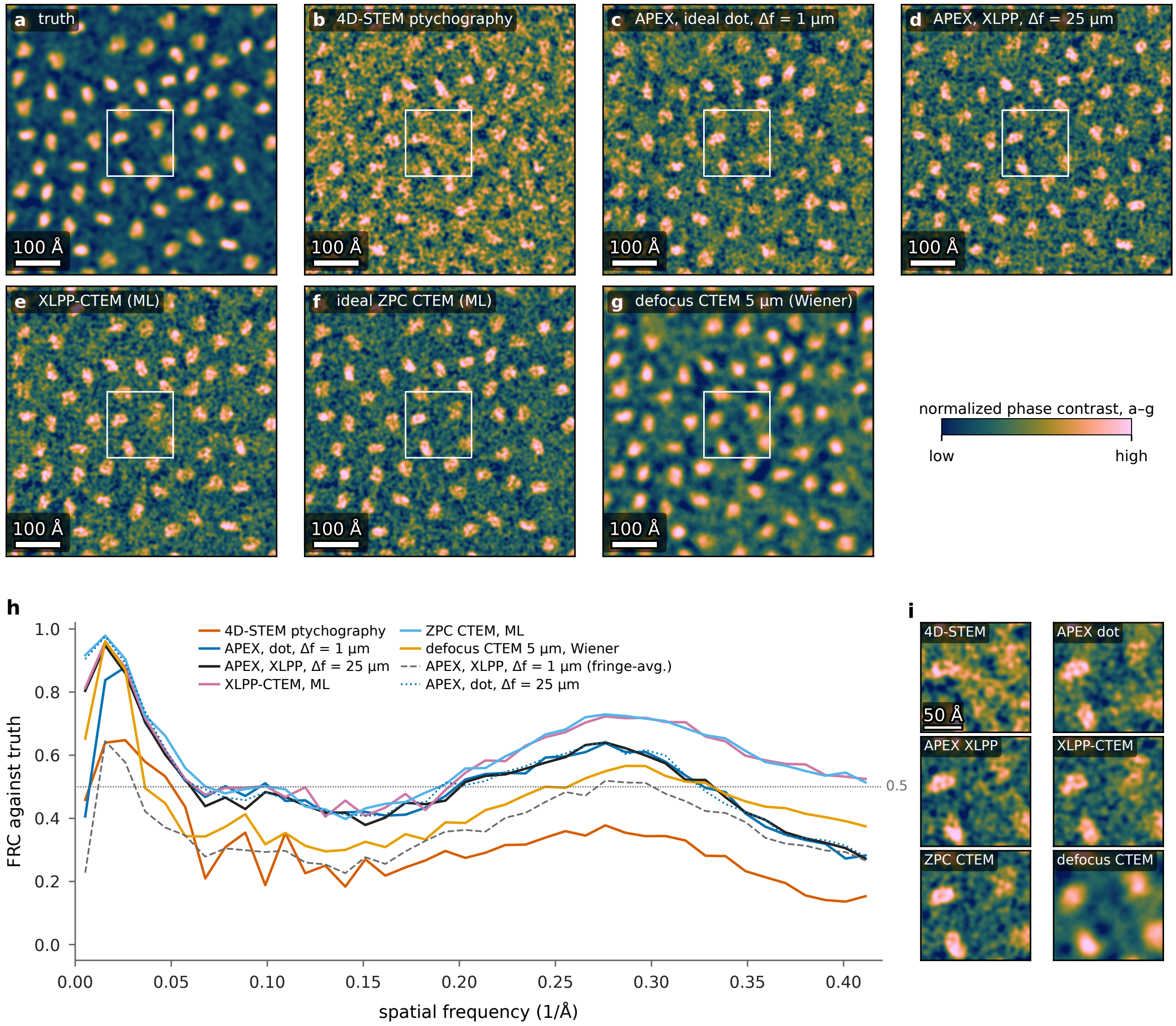

FIG. 5. Two-dimensional phase reconstructions of solvated myoglobin in 60 Å of vitreous ice at 30 e/Å² (300 keV). (a) Ground-truth projected phase. (b) Defocused 4D-STEM ptychography. (c) APEX with an ideal $\frac{\pi}{2}$ dot ($\Delta f = 1$ µm). (d) APEX with an XLPP ($\Delta f = 25.4$ µm). (e) Plane-wave CTEM with an XLPP in the back focal plane. (f) Ideal ZPC CTEM. (g) Defocused plane-wave CTEM. The colour scale shows normalized phase contrast, not phase in radians. (h) Fourier ring correlation (FRC) against the ground-truth phase. Dashed grey: APEX with the XLPP at $\Delta f = 1$ µm; dotted: APEX with the ideal dot at $\Delta f = 25.4$ µm. (i) Enlarged views of the region marked by the white box in (a)–(g). Supplementary Note S2 gives the display filters and the reconstruction settings.

The results at $\Delta f = 25.4$ µm hold for a periodic specimen model. We simulated APEX with the XLPP (38 kW per cavity, Sec. III) at two defoci, because the unscattered spot at $C'$ receives only about half of the $\frac{\pi}{2}$ phase at $\Delta f = 1$ µm, but nearly all of it inside a single antinode at $\Delta f =$ 25.4 µm (Sec. IV B, Table S1). At the larger defocus, the illuminated patch is larger than the specimen cell, so we tiled the periodic specimen and used a coarse scan. Each frame then illuminates several copies of the same cell, and the dose of all copies falls onto one periodic unknown: the 30 e/Å² of these runs is the dose folded onto the cell, not the local fluence on each distinct particle of a non-repeating specimen. The reconstruction (a full-batch variant of the forward model with modulo indexing, explicitly band-limited to the objective aperture) uses this periodicity, which also makes the content of the Kapitza–Dirac ghosts known. We did not test a non-repeating specimen in two dimensions at this defocus. In the depth benchmark at $\Delta f = 9.3$ µm, however, non-repeating mosaic screens gave the same scores as the periodic cell at matched local fluence (Sec. VI B). Single-scan depth sectioning at $\Delta f = 25.4$ µm would further require a scan covering the full patch diameter $2\alpha\Delta f$, so that every specimen point is illuminated by the full cone of incidence angles; we did not simulate it.

As shown in Fig. 5 (h) and Fig. 6, APEX with the ideal dot at $\Delta f = 1$ µm outperforms defocused 4D-STEM ptychography at 300 keV in every metric of Table S5. It is also higher in all sampled FRC bins except the lowest one, below the cut-on of the dot. The FRC against the ground-truth phase more than doubles at 0.1 Å⁻¹, and the median per-particle correlation rises markedly (Table S5). Because unregularized ML iterations do not attenuate noisy high-frequency shells, we additionally applied an identical, truth-free radial SNR filter to every ML reconstruction, weighting each Fourier shell by $2\frac{F_h}{1+F_h}$ obtained from a binomial half-dose split. This isotropic weight leaves the shell-wise FRC against the ground truth nearly unchanged but suppresses high-frequency noise in real space; it benefits 4D-STEM most, yet APEX retains a wide lead (Fig. 6 b).

The probe defocus $\Delta f$ determines how closely APEX approaches plane-wave phase-plate CTEM (Fig. 6, Table S5). At $\Delta f = 1$ µm, the ideal dot stays below ZPC CTEM in all scores. At $\Delta f = 25.4$ µm, APEX with the dot matches the per-particle value of ZPC CTEM after SNR weighting and nearly matches its correlation at the finest low-pass radius, while its FRC at high spatial frequencies remains lower. The realistic XLPP shows the same trend more strongly. At $\Delta f = 1$ µm, where its spot receives only the fringe-averaged phase, APEX with the 38 kW plate reaches only 0.56 per particle, well below the ideal dot (Table S5); at the larger defocus, the same plate comes close to XLPP-CTEM. In one control realization at $\Delta f = 50.4$ µm, the scores did not change (Table S3). In the periodic model, a defocus of about 25 µm therefore appears sufficient, and the remaining per-particle gap between the XLPP and the dot does not close with further defocus. At large defocus, each APEX frame is effectively a phase-plate image of a large patch, such that, in the periodic model, APEX approaches the two-dimensional performance of phase-plate CTEM at low and medium resolution. We simulated depth sectioning with the XLPP at a smaller defocus and a larger convergence angle (Sec. VI B), not at $\Delta f = 25.4$ µm. A defocused CTEM image at 5 µm, reconstructed with a Wiener filter, exceeds APEX at $\Delta f = 1$ µm in all low-pass correlations, but its FRC is lower than that of ideal-dot APEX, CTEM with the XLPP and APEX with the large-defocus XLPP (Fig. 5 h). The solvers of these rows differ, so this comparison does not isolate the phase element. It suggests that the advantage of a well-matched phase element lies in the FRC transfer rather than in the low-pass scores.

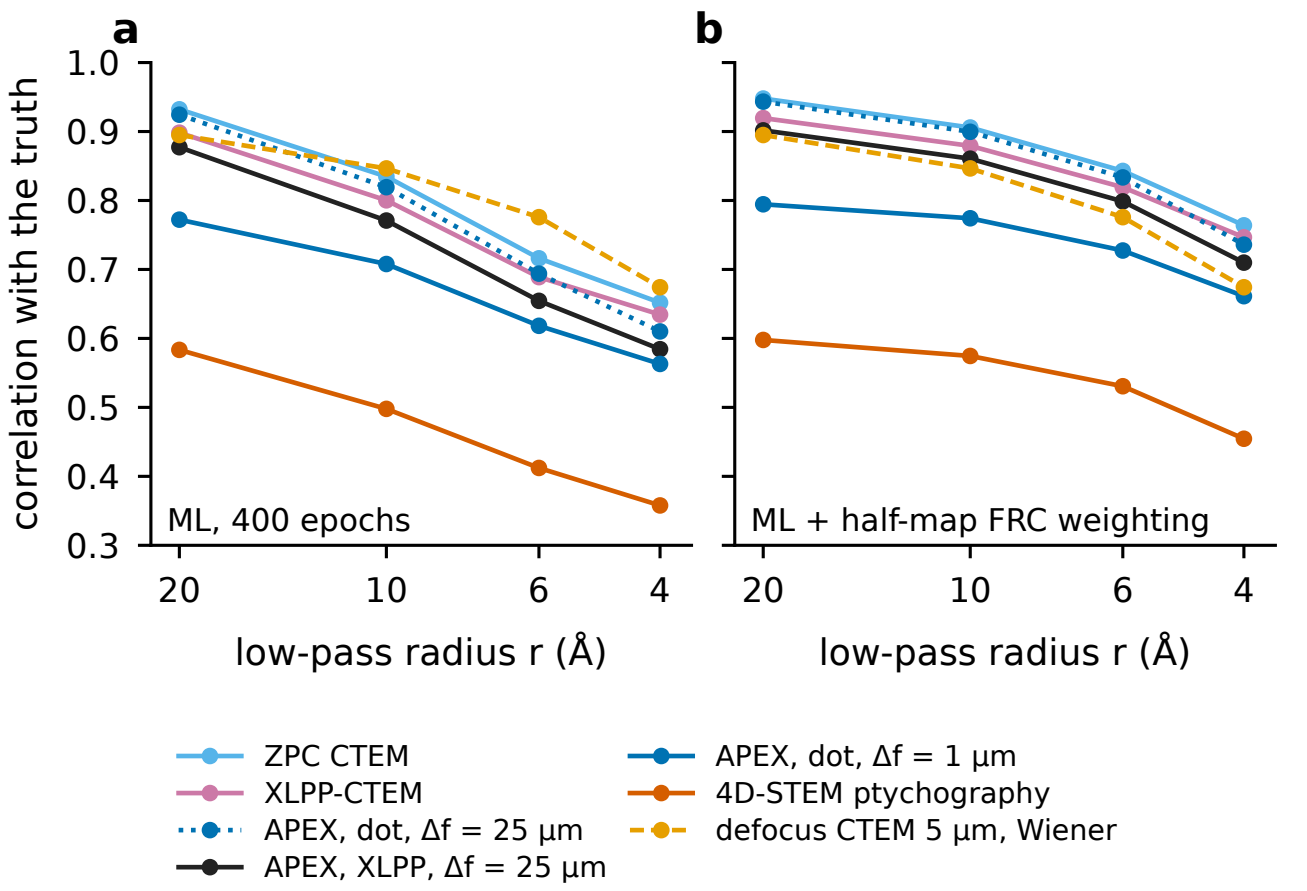

FIG. 6. Correlation of the myoglobin phase reconstructions of Fig. 5 with the ground-truth phase after a Gaussian low-pass at radius $r$. (a) Poisson ML. (b) The same reconstructions with the half-map FRC weight described in the text. Table S5 lists the values; Supplementary Note S2 gives the filter.

The three Poisson realizations, a fit of the phase scale and a longer iteration budget support these rankings (Table S3). The gap between APEX with the XLPP and XLPP-CTEM exceeds the noise spread at high spatial frequencies, whereas APEX with the dot at $\Delta f = 25.4$ µm and ZPC CTEM agree per particle within the noise spread. Because correlations do not see the phase scale, we also fitted each reconstruction to the truth after a 10 Å low-pass. APEX and CTEM recover the phase scale to within about 15%, the largest deficit occurring with the dot at $\Delta f = 1$ µm, where its cut-on suppresses the lowest frequencies. 4D-STEM recovers the scale, but its rms phase error exceeds the rms of the filtered truth. A longer iteration budget narrows the per-particle gap between APEX with the XLPP and XLPP-CTEM to the noise spread, but it leaves the high-frequency gap unchanged, so this gap does not depend on the iteration count. For 4D-STEM, a sweep of learning rate and epochs improves the per-particle value by at most 0.04 over the setting used here. Any physical CTEM phase plate has a finite cut-on that reduces the low-frequency transfer relative to the ideal Zernike row.

## V. PHASE RECONSTRUCTION

We reconstruct the specimen phase by fitting a physical model of every recorded frame to the measured counts (Fig. S1). For the two-dimensional micrographs, the only unknown is a single phase screen $\varphi(\boldsymbol{r})$, sampled on $N_o$ object pixels; the incident probe, the $\frac{\pi}{2}$ element and the defocus $\Delta f$ are known from the optics and held fixed. The expected counts of the frame at scan position $\boldsymbol{p}$ follow the post-specimen chain of Sec. II C:

$$\mu_p(\boldsymbol{r}) = n_e \Big| \mathcal{P}_{+\Delta f}\big\{ t(\boldsymbol{r}-\boldsymbol{c}_p)\,\mathcal{P}_{-\Delta f} \times \mathcal{A}\big[\psi_0(\boldsymbol{r}-\boldsymbol{p})e^{i\varphi(\boldsymbol{r})}\big]\big\}\Big|^2, \tag{4}$$

where $n_e$ is the number of electrons per frame, $\mathcal{A}$ is the objective aperture, $\mathcal{P}_{\pm\Delta f}$ is Fresnel propagation to the crossover image $C'$ and back to the specimen image $S'$, and $t$ is the transmission of the phase element, centred on the unscattered spot at $\boldsymbol{c}_p$, which follows the scan (Fig. S1 a–c). Each of the $N_p$ frames has $N_r$ detector pixels $\boldsymbol{r}$, so that $\mu_p$ and the measured counts $I_p$ are arrays of $N_r$ values per frame. For the measured counts $I_p(\boldsymbol{r})$ (Fig. S1 d), we minimize the Poisson negative log-likelihood

$$L(\varphi) = \sum_{p,\boldsymbol{r}} \big[\mu_p(\boldsymbol{r}) - I_p(\boldsymbol{r}) \ln \mu_p(\boldsymbol{r})\big] \tag{5}$$

with the Adam optimizer. The gradient $\partial L/\partial\varphi$ follows from automatic differentiation through Eq. (4), and each step uses a minibatch of neighbouring scan positions (Table S1). After each step, the object is projected onto a pure-phase object.

The loss decreased in all our simulations. Two properties of the data help the fit. First, in the weak-phase regime each frame is already a ZPC image of its illuminated patch, $\mu_p(\boldsymbol{x}) \approx n_e\ |\psi_{0,p}(\boldsymbol{x})|^2\ \big[1 + 2\sigma(\varphi(\boldsymbol{x}) - \langle\varphi\rangle_p)\big]$ above the cut-on $q_c$ (Fig. S1 e), so that a normalized weighted sum over scan positions already gives a high-pass estimate of the phase (analogous to Eq. 36 of Dwyer and Paganin[9]). To first order in $\varphi$, the gradient of Eq. (5) at $\varphi = 0$ is also a weighted sum of these contrast images over scan positions. Second, the patches of neighbouring positions overlap: with the scan step of Table S1, each specimen point appears in about ten frames, each time at a different position in the patch and therefore under a different incidence angle. This redundancy helps to constrain the low frequencies near the cut-on, which a single frame transfers poorly. In principle, it also allows the probe, the element transmission and $\Delta f$ to be refined together with the object (the calibration step left open for far-field X-ray data in Vartiainen *et al.*[13]); we did not test this joint refinement here. For the myoglobin micrographs, the correlation with the true phase continues to rise between 100 and 400 epochs (Fig. S1 g, h).

We implemented this image-plane detector stage in the `scatterem` package. The myoglobin reconstructions in Fig. 5 and Fig. 6 use it with the probe and the phase element held fixed (the $\Delta f = 25.4$ µm rows use its periodic-cell variant, Sec. IV E). For depth sectioning (Sec. VI A), we use the same chain with a multislice object, linearized once at $\varphi = 0$, and solve for all layers in one linear step (Fig. S1 i–l).

## VI. THREE-DIMENSIONAL DEPTH SECTIONING

Whereas a single plane-wave CTEM exposure projects the entire specimen thickness along a single illumination direction and therefore carries no depth discrimination beyond Ewald-sphere curvature, a defocused APEX probe illuminates each specimen coordinate $\boldsymbol{r}$ from an incidence angle $\boldsymbol{\theta} = \frac{\boldsymbol{r}-\boldsymbol{p}}{\Delta f}$ that sweeps continuously across the convergence cone as the probe position $\boldsymbol{p}$ scans laterally. Consequently, a single two-dimensional APEX scan records a continuous, limited-angle tilt series with ZPC.

### A. Point-projection geometry and linear multi-layer reconstruction

Because the incident probe approximates a spherical wave diverging from the crossover $C$, the Fresnel scaling theorem[31] implies that a specimen phase layer $\varphi_z$ located at axial depth $z$ below the in-focus conjugate plane appears in frame $p$ as a point projection magnified about the scan position $\boldsymbol{p}$:

$$L_p(\boldsymbol{r}) = 2\sigma\ |B_p(\boldsymbol{r})|^2 \sum_z \mathrm{Re}\Big\{P_{d_z}\varphi_z\Big\}(\boldsymbol{x}_z), \qquad \boldsymbol{x}_z = \boldsymbol{p} + \frac{\boldsymbol{r}-\boldsymbol{p}}{M_z}, \quad M_z = \frac{\Delta f}{\Delta f + z}, \tag{6}$$

where $B_p$ is the incident probe amplitude in the focus plane and $P_{d_z}$ denotes Fresnel propagation over the effective defocus distance $d_z = -z\frac{\Delta f + z}{\Delta f}$. Each recorded frame thus superposes ZPC images of all axial layers shifted laterally by the parallax offset $z\boldsymbol{\theta}$. Eq. (6) is a geometric approximation that holds above the cut-on of the phase element. It omits the reference projection of the exact Zernike operator (Sec. II C), which subtracts from each layer its probe-weighted mean phase, as in the thin-object expression of Sec. V: a spatially constant layer changes only the global phase and gives no signal, whereas Eq. (6) would predict $2\sigma\ |B_p|^2\ c$ for $\varphi_z = c$. The numerical solver below does not use Eq. (6); it uses the full Jacobian of the finite-dot forward model, which contains this subtraction, and we checked that it gives no response to a constant layer. From this forward geometry, we derive two linear depth-sectioning algorithms, analogous to tilt-corrected bright-field STEM[32] and linearized multislice ptychography[33]:

1. **Magnify-and-add synthetic refocusing.** Resampling every frame $p$ about its probe centre $\boldsymbol{p}$ by the magnification factor $M_{z_0}$, normalizing by the resampled probe intensity $2\,|B_p|^2$, and summing across all scan positions $p$ brings the target plane $z_0$ into coherent registration while blurring any layer separated by $\Delta z$ over a disk of radius $\alpha\,|\Delta z|$ (reflecting the missing cone of a $\pm\alpha$ tilt series).
2. **Joint linear Wiener inversion.** Linearizing the exact multislice[33] and post-specimen forward model around $\varphi = 0$, weighting the data residuals by the expected Poisson shot-noise variance, incorporating a stationary Gaussian spatial prior, and solving jointly for all depth layers simultaneously (Eq. (7) and Eq. (9)) deconvolves inter-layer cross-talk and contrast-transfer modulations in a single step.

To formulate the joint linear Wiener solver explicitly, we stack the data and the unknowns into vectors. The data vector $\boldsymbol{I} \in \mathbb{R}^{N_p N_r}$ holds the measured detector counts of all $N_p$ frames $p$ and all $N_r$ pixels $\boldsymbol{r}$ per frame, and $\boldsymbol{\mu}^{(0)} \in \mathbb{R}^{N_p N_r}$ holds the expected counts of the unperturbed reference frames at $\varphi = 0$. The unknown $\boldsymbol{\varphi} = \left(\varphi_{z_1}, \ldots, \varphi_{z_{N_z}}\right) \in \mathbb{R}^{N_z N_o}$ stacks the $N_z$ phase layers, each sampled on $N_o$ object pixels. The forward-model Jacobian $\boldsymbol{J} = \partial\boldsymbol{\mu}/\partial\boldsymbol{\varphi} \in \mathbb{R}^{N_p N_r \times N_z N_o}$ is evaluated at $\varphi = 0$. We use a Gaussian surrogate of the Poisson likelihood: the linearized intensities with fixed reference-state Poisson variances and a floor for dark pixels. We minimize the maximum *a posteriori* estimate regularized by a Gaussian prior with spatial power spectrum $S(q)$:

$$
\begin{aligned}
\hat{\boldsymbol{\varphi}} = \arg\min_{\boldsymbol{\varphi}} \frac{1}{2} \sum_{p,\boldsymbol{r}} \frac{\left(I_p(\boldsymbol{r}) - \mu_p^{(0)}(\boldsymbol{r}) - [\boldsymbol{J}\boldsymbol{\varphi}]_p(\boldsymbol{r})\right)^2}{v_p(\boldsymbol{r})} \\
+ \frac{1}{2} \sum_z \sum_q \frac{|\Phi_z(q)|^2}{S(q)},
\end{aligned} \tag{7}
$$

where $v_p(\boldsymbol{r}) = \max\left(\mu_p^{(0)}(\boldsymbol{r}), \varepsilon\right)$ with relative variance floor $\varepsilon = 10^{-3} \max \mu^{(0)}$ for dark pixels, and $\Phi_z = \mathcal{F}\varphi_z$. Minimizing Eq. (7) yields the normal equations

$$
(\boldsymbol{J}^T \boldsymbol{V}^{-1} \boldsymbol{J} + \boldsymbol{\Sigma}^{-1})\hat{\boldsymbol{\varphi}} = \boldsymbol{J}^T \boldsymbol{V}^{-1}\left(\boldsymbol{I} - \boldsymbol{\mu}^{(0)}\right), \tag{8}
$$

with the diagonal noise covariance $\boldsymbol{V} = \mathrm{diag}(v) \in \mathbb{R}^{N_p N_r \times N_p N_r}$ and the prior covariance $\boldsymbol{\Sigma} \in \mathbb{R}^{N_z N_o \times N_z N_o}$. $\boldsymbol{\Sigma}$ is block diagonal with one $N_o \times N_o$ block per layer, and each block is diagonal in Fourier space with entries $S(q)$. Both sides of Eq. (8) are vectors in $\mathbb{R}^{N_z N_o}$. Transforming to prior-whitened variables $\hat{\boldsymbol{\varphi}} = \boldsymbol{\Sigma}^{\frac{1}{2}}\boldsymbol{\psi}$, $\boldsymbol{\psi} \in \mathbb{R}^{N_z N_o}$, yields the well-conditioned $N_z N_o \times N_z N_o$ system matrix $\mathbf{1} + \boldsymbol{\Sigma}^{\frac{1}{2}} \boldsymbol{J}^T \boldsymbol{V}^{-1} \boldsymbol{J} \boldsymbol{\Sigma}^{\frac{1}{2}}$. We never form this matrix: we solve the system by matrix-free conjugate gradients, with one Jacobian-vector and one vector-Jacobian product per iteration evaluated by automatic differentiation in pytorch.

In the limit of a dense scan and uniform noise weights, Eq. (8) decouples across lateral spatial frequencies. Taking the two-dimensional Fourier transform of the linearized intensity modulation per electron $L_p(\boldsymbol{r}) = \left[\mu_p(\boldsymbol{r}) - \mu_p^{(0)}(\boldsymbol{r})\right]/n_e$ with respect to image coordinate $\boldsymbol{r}$ (conjugate frequency $\boldsymbol{u}$) and scan position $\boldsymbol{p}$ (conjugate frequency $\boldsymbol{Q}$), and identifying the specimen spatial frequency $\boldsymbol{q} = \boldsymbol{u} + \boldsymbol{Q}$, the forward model for an ideal phase dot takes the form

$$
\begin{aligned}
\tilde{L}(\boldsymbol{u}, \boldsymbol{Q}) &= \sigma \sum_z H_z(\boldsymbol{u}, \boldsymbol{q}) \Phi_z(\boldsymbol{q}), \\
H_z(\boldsymbol{u}, \boldsymbol{q}) &= K_z(\boldsymbol{u}, \boldsymbol{q}) + K_z^*(-\boldsymbol{u}, -\boldsymbol{q}),
\end{aligned} \tag{9}
$$

$$
\begin{aligned}
K_z(\boldsymbol{u}, \boldsymbol{q}) &= \int a(\boldsymbol{k} - \boldsymbol{u}) a(\boldsymbol{k} - \boldsymbol{q}) e^{i\Theta}\, \mathrm{d}\boldsymbol{k}, \\
\Theta &= \pi\lambda\left[2(\Delta f \boldsymbol{Q} + z\boldsymbol{q}) \cdot \boldsymbol{k} + \Delta f(u^2 - q^2) - zq^2\right],
\end{aligned}
$$

where $a$ is the condenser aperture indicator function. Like Eq. (6), this ideal-dot response holds above the cut-on: it omits the reference projection, and at $\boldsymbol{q} = \boldsymbol{u} = \boldsymbol{Q} = 0$ it gives $H_z = 2\int a^2\, \mathrm{d}\boldsymbol{k}$ instead of zero, so the modes below the cut-on are excluded. Because the integral over the overlap of the two shifted aperture disks concentrates around stationary phase where $\Delta f \boldsymbol{Q} + z\boldsymbol{q} = 0$, each depth plane $z$ maps onto a tilted plane $\boldsymbol{Q} = -z\frac{\boldsymbol{u}}{\Delta f + z}$ in the four-dimensional Wigner spectrum, directly analogous to Fourier-slice light-field refocusing, which Ma *et al.* adapted to tilt-corrected bright-field depth sectioning in 4D-STEM[34]. At each lateral spatial frequency $\boldsymbol{q}$, Eq. (8) then reduces to a compact $N_z \times N_z$ linear system that couples the depth layers. With the noise power $N$, the prior power spectrum $S(q)$ of Eq. (7) and the Fourier transform $\Phi_z(\boldsymbol{q})$ of phase layer $z$, it reads

$$
\begin{aligned}
\sum_{z'} G_{zz'}(\boldsymbol{q}) \Phi_{z'}(\boldsymbol{q}) &= \sigma \sum_{\boldsymbol{u}} H_z^*(\boldsymbol{u}, \boldsymbol{q}) \tilde{L}(\boldsymbol{u}, \boldsymbol{q} - \boldsymbol{u}), \\
G_{zz'}(\boldsymbol{q}) &= \sigma^2 \sum_{\boldsymbol{u}} H_z^* H_{z'} + \delta_{zz'} \frac{N}{S(q)},
\end{aligned}
$$

This system is the multi-layer generalization of single-sideband ptychography. With $\varphi$ in radians, as in our implementation, $\sigma = 1$. Solving Eq. (8) via conjugate gradients generalizes this deconvolution to arbitrary finite scan steps, reference-state Poisson pixel weights and realistic laser or point-charge phase elements.

Globally coupled linear ptychographic solvers have previously been formulated for two-dimensional objects in far-field diffraction: D'Alfonso *et al.* linearized the diffraction-pattern autocorrelations across all probe positions and solved them by conjugate-gradient least squares with early stopping[35–37], and Yan *et al.* relinearized around the current object estimate in X-ray ptychography[38]. Our solver extends these linear formulations in four respects: (i) it models multiple axial specimen layers through a multislice forward operator; (ii) it linearizes the measured detector intensities directly with reference-state Poisson variance weights (a Gaussian surrogate of the Poisson likelihood) rather than unweighted pattern autocorrelations; (iii) it incorporates an explicit spectral Wiener prior in place of early-stopping regularization; and (iv) it applies identically to conjugate image-plane (APEX) and diffraction-plane (4D-STEM) datasets. Furthermore, whereas the multislice electron ptychography scheme of Bangun *et al.* updates one axial slice at a time while holding all other slices fixed[33], our joint solver inverts the cross-talk matrix $G_{zz'}$ across all layers simultaneously in a single step.

From the geometric blur disk of radius $\alpha\,|\Delta z|$, the half-width at half-maximum of the depth response is $\frac{0.35}{\alpha q}$; we call it the geometric axial half-width. We compared this estimate with a noise-free simulation of magnify-and-add refocusing, in which each of two random phase layers is propagated alone and the reference wave receives an ideal phase shift (Fig. 7 (h), Table S4). For the layer in the focus plane, the response peaks at the true depth with a half-width close to $\frac{0.35}{\alpha q}$ in all three frequency bands. For the displaced layer, the two lower bands behave in the same way, whereas the highest band gives a weak response that peaks short of the true depth. Magnify-and-add does not correct the Fresnel propagation of a displaced layer, whose transfer $\cos\left(\pi\lambda d_z q^2\right)$ approaches its first zero in this band. The value $\frac{0.35}{\alpha q}$ is therefore a geometric blur width, not the axial localization that a reconstruction recovers from noisy data. This $\alpha$-limited missing cone is the same as the one Terzoudis-Lumsden *et al.* derived for scattering-matrix parallax reconstruction from 4D-STEM, which they showed to be equivalent to scanning confocal imaging with a hypothetical phase-sensitive point detector[39]; in APEX, the Zernike element provides this phase sensitivity optically, so it needs no prior retrieval of the scattering matrix.

### B. Depth sectioning of two myoglobin phase screens 150 nm apart

To benchmark depth sectioning quantitatively at cryo-EM fluences, we constructed a two-phase-screen phantom. Each screen is the projected phase of a region of our atomistic myoglobin field (two non-overlapping regions), and the screens are placed 150 nm apart, at $z = \mp 750$ Å (Table S1). The wave is multiplied by the first screen, propagated through 150 nm of vacuum and multiplied by the second screen; the phantom contains no ice between the screens, no absorption and periodic lateral boundaries, and is therefore not a model of a continuous 150 nm cryogenic lamella. We simulated this propagation with Poisson shot noise for APEX, 4D-STEM with the identical incident probe (the convergence semi-angle of the two-dimensional study, crossover 1 μm above the central focus plane, and an ideal $\frac{\pi}{2}$ dot of radius 2 Airy units for APEX), and plane-wave ZPC CTEM at matched total fluences of 30 e/Å² and 100 e/Å². All three modalities were inverted using the same joint linear Wiener solver and the same isotropic spectral prior (calibrated from the power spectrum of a third independent myoglobin region), with the reconstruction restricted to the two known screen depths, and the scores were averaged across three independent Poisson noise seeds (Fig. 7 and Table I). This benchmark therefore tests the recovery of two prescribed layer maps; it does not test the localization of layers at unknown depths, an unknown number of layers, or a continuous three-dimensional specimen.

As expected from projection geometry, a single ZPC CTEM exposure cannot separate the two axial planes: both reconstructed layers are identical mixtures that correlate equally with their own layer and with the opposite layer (Fig. 7 (d), Table I). Both APEX and 4D-STEM ptychography separate the two depths above 0.06 Å⁻¹, where the cross-talk is negligible. Below this frequency, about a third of each layer leaks into the other at the lower fluence, for APEX and 4D-STEM alike (Table I). This is the band in which the missing cone becomes relevant: there, the geometric axial half-width $\frac{0.35}{\alpha q}$ exceeds the screen spacing (Table S4). Within the two-screen model with known depths, however, the geometry alone does not prevent the separation, and the cross-talk is set by noise: at a very high fluence, it nearly vanishes (Table S4), so that at low dose the noise and the prior determine how the solver distributes the lowest frequencies between the screens. This does not show that the missing cone loses no information for a continuous specimen or for screens at unknown depths. After the 10 Å low-pass, the joint Wiener APEX reconstruction recovers each screen better than 4D-STEM at both fluences (Table I). Within the highest band, however, 4D-STEM slightly exceeds APEX. A likely cause, which we did not test separately, is that the out-of-focus layers encounter the first zero of their Fresnel propagation transfer $\cos\left(\pi\lambda d_z q^2\right)$ at $q = \left(2\lambda\,|d_z|\right)^{-\frac{1}{2}}$: image-plane APEX measures a single quadrature of the defocused wave at each layer, whereas the interfering diffraction disks of 4D-STEM retain both quadratures. In the 1 μm benchmark, the joint Wiener solver outperforms simple magnify-and-add refocusing (Table I), because it deconvolves this defocus transfer and normalizes the spatially varying probe weights. Damped Gauss–Newton relinearization steps around the current estimate (analogous to Yan *et al.*[38]), which minimize the Poisson objective and so include the quadratic dark-field signal, leave APEX unchanged and raise 4D-STEM only slightly, still below the linear APEX reconstruction (Table S4); we have not established the ordering of a fully converged nonlinear reconstruction.

We repeated the benchmark at large defocus with the realistic 38 kW XLPP of Fig. 3 at its full fringe structure (at the $\alpha\Delta f$ of Sec. IV D), the ideal dot as a control, and plane-wave ZPC CTEM, with the same solver, prior, screen depths and fluences. Each screen is a non-repeating mosaic of tiles of the myoglobin cell, where every tile has its own random rotation, reflection and cyclic shift, and the two screens use independent draws (Table S1). The scan covers the periodic window so that every point is seen from many directions. The XLPP recovers each screen far better than APEX at $\Delta f = 1$ µm, close to the ideal dot, with negligible cross-talk in every band (Table I). The larger aperture also removes the low-frequency cross-talk of the 1 µm case: in the lowest band, the geometric axial half-width $0.35/(\alpha q)$ is well below the screen spacing (Table S4). A run on the periodic cell, where the dose of all illuminated copies falls onto one unknown, gives the same scores (Table S4); at matched local fluence, the periodic model therefore does not inflate the scores. One ZPC CTEM exposure again cannot separate the screens.

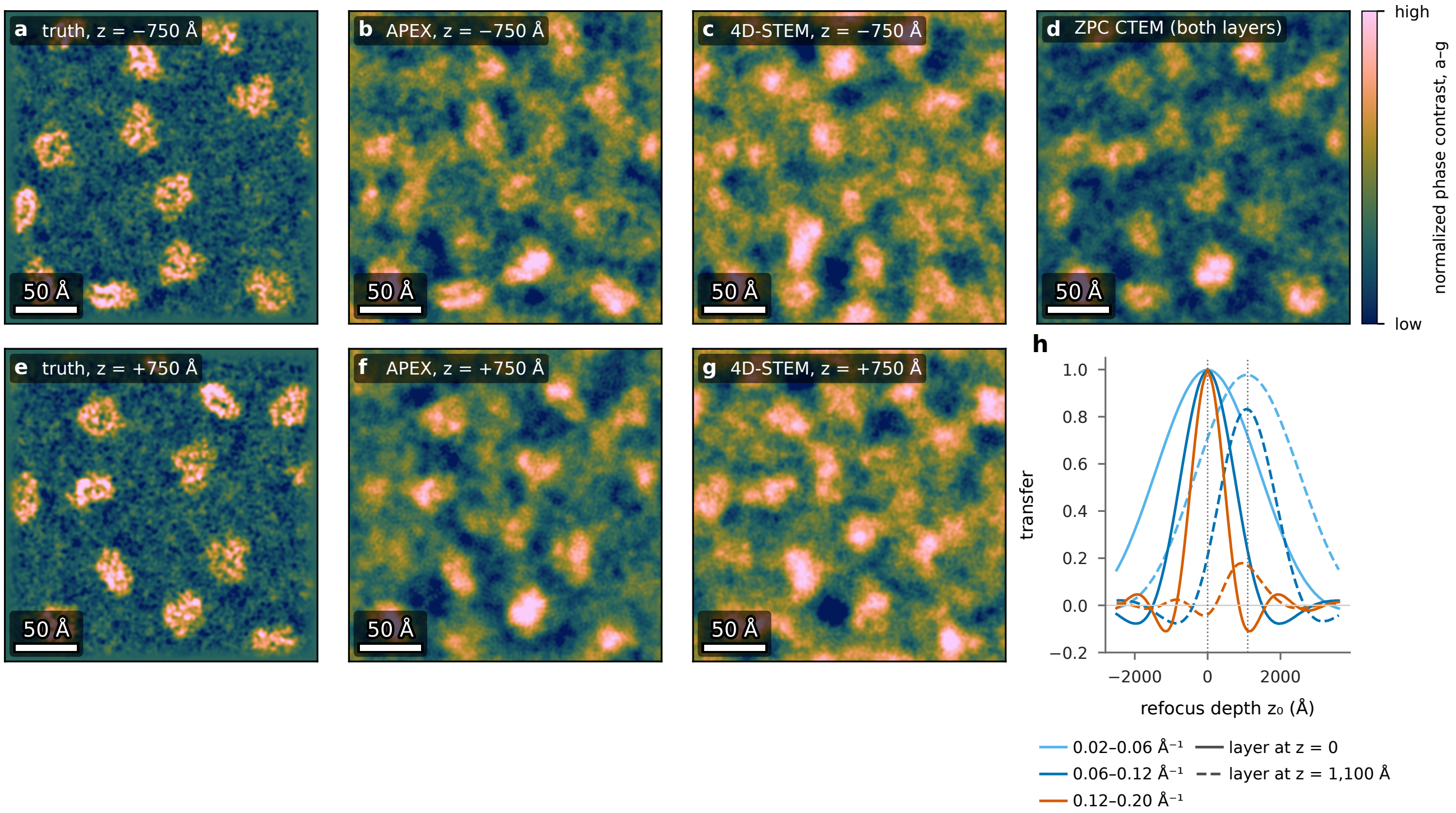

FIG. 7. Depth sectioning of two myoglobin phase screens 150 nm apart from a single scan at 30 e/Å². (a–c) Upper layer: ground truth, APEX and 4D-STEM. (d) Plane-wave ZPC CTEM: both reconstructed layers collapse to the same superposition of the two true layers. (e–g) The same for the lower layer. (h) Noise-free axial depth-response profiles of magnify-and-add refocusing for a layer at $z = 0$ (solid) and at $z = 1,100$ Å (dashed), in three spatial-frequency bands. Supplementary Note S2 gives the parameters.

TABLE I. Depth sectioning of the two-screen phantom of Sec. VI B: correlation of each reconstructed layer with its own and with the opposite true layer (cross-talk), given as layer 1 / layer 2. Shading from 0 (white) to 1; cross-talk in red. The tilt series are idealized (see text). Supplementary Note S2 gives the filters, the seeds and the standard deviations.

| Mode | 30 e/Å² (10 Å low-pass) | | 30 e/Å², 0.02–0.06 Å⁻¹ | 100 e/Å² (10 Å) | 30 e/Å², 0.12–0.2 Å⁻¹ |
|---|---|---|---|---|---|
| | own layer | other layer | other layer | own layer | own layer |
| **$\Delta f = 1$ µm, $\alpha = 5.47$ mrad (CTEM: plane wave)** | | | | | |
| APEX, Wiener solve | 0.64 / 0.67 | 0.30 / 0.32 | 0.35 / 0.29 | 0.79 / 0.83 | 0.24 / 0.21 |
| ZPC CTEM, Wiener solve | 0.65 / 0.64 | 0.64 / 0.65 | 0.65 / 0.64 | 0.69 / 0.67 | 0.16 / 0.17 |
| 4D-STEM, Wiener solve | 0.49 / 0.54 | 0.33 / 0.33 | 0.34 / 0.33 | 0.64 / 0.68 | 0.26 / 0.27 |
| APEX, magnify-and-add | 0.37 / 0.42 | 0.19 / 0.20 | 0.44 / 0.39 | 0.49 / 0.55 | 0.16 / 0.04 |
| **$\Delta f = 9.3$ µm, $\alpha = 15$ mrad, non-repeating mosaic layers (CTEM: plane wave)** | | | | | |
| APEX, ideal dot, Wiener solve | 0.86 / 0.85 | 0.03 / 0.03 | 0.00 / 0.01 | – | 0.23 / 0.20 |
| APEX, XLPP 38 kW, Wiener solve | 0.83 / 0.82 | 0.05 / 0.05 | 0.02 / 0.02 | 0.91 / 0.90 | 0.25 / 0.20 |
| 4D-STEM, Wiener solve | 0.78 / 0.78 | 0.03 / 0.03 | 0.00 / 0.01 | – | 0.32 / 0.32 |
| ZPC CTEM, one exposure, Wiener solve | 0.64 / 0.64 | 0.64 / 0.64 | 0.63 / 0.63 | 0.68 / 0.68 | 0.18 / 0.17 |
| ZPC CTEM, 41 tilts to ±60°, Wiener solve | 0.87 / 0.87 | 0.00 / 0.00 | 0.01 / 0.01 | 0.93 / 0.93 | 0.26 / 0.26 |
| ZPC CTEM, 29 beam tilts within $\alpha$, Wiener solve | 0.86 / 0.86 | 0.02 / 0.02 | 0.00 / 0.00 | 0.93 / 0.93 | 0.19 / 0.18 |

4D-STEM with the same probe and scan also separates the screens (Table I). After the 10 Å low-pass, it recovers each screen less well than APEX. In the 0.12–0.2 Å$^{-1}$ band it is higher than APEX, as in the 1 µm benchmark. We also simulated two series of ZPC CTEM images at the same total dose. The first series has 41 specimen tilts up to ±60°. The second series has 29 beam tilts inside the convergence cone of APEX. The model of both series is idealized: a tilt shifts each screen laterally in proportion to its height, and every image has the full Zernike transfer. The model does not include the foreshortening, the longer path in the ice or the alignment of the series. It is therefore an upper bound for a real tilt series. Both series separate the screens as well as APEX with the ideal dot (Table I). This agrees with Sec. VI: one APEX scan is a limited-angle tilt series. In this benchmark, APEX therefore does not separate the screens better than a tilt series of ZPC CTEM. APEX records the directions in one scan of an untilted specimen. The beam-tilt series needs the unscattered beam to stay on the phase element at every tilt.

These results suggest two design rules (Table S4). First, because the effective cut-on frequency of the phase dot depends on layer depth as $q_c(z) \approx \frac{\rho}{\lambda(\Delta f + z)}$, where $\rho$ is the radius of the phase dot, specimen layers located too close to the probe crossover lose low-frequency contrast. In one comparison, a crossover only half as far above the mid-plane lowered the score of the upper layer, for APEX and for 4D-STEM with the same probe (Table S4). The cut-on of the dot is therefore not the only cause. Until a further comparison shows otherwise, we place the probe crossover well upstream of the specimen entrance surface. Second, the geometric axial half-width $\frac{0.35}{\alpha q}$ scales inversely with the lateral spatial frequency $q$. Separating whole protein envelopes from a single scan at the $\alpha$ of Fig. 7 therefore requires axial layer separations of several hundred nanometres. A larger $\alpha$ relaxes this requirement: at $\alpha = 15$ mrad, the cross-talk of APEX on the same phantom remained small and became large only when the screen spacing was reduced to a third (Table S4). A larger $\alpha$, however, also shrinks the crossover Airy disk that the element must resolve (Sec. IV D).

## VII. DISCUSSION AND OUTLOOK

In the weak-phase regime, the post-specimen optics form the phase contrast directly on the detector in each exposure, such that each exposure supplies local phase contrast above the cut-on without an overlap-based phase-retrieval algorithm; consistent phase offsets between separate illuminated patches may still require overlap or additional constraints. In its linear form, APEX is therefore a scanned ZPC imaging mode rather than ptychography in the strict sense of Hoppe or Rodenburg; the ptychographic overlap constraint comes into play when jointly refining the complex probe, the phase-element transmission and multi-slice or non-linear object phases. Vartiainen *et al.* likewise relied on the overlap of far-field diffraction patterns in their ptychographic reconstruction of Zernike X-ray data[13].

In our simulations at large probe defocus, APEX with a realistic XLPP comes close to the two-dimensional performance of phase-plate CTEM, with a remaining gap at high spatial frequencies. With the same plate at a smaller defocus and a larger convergence angle, a single scan also separates two non-repeating phase screens at known depths, nearly as well as with an ideal phase dot (Table I). This depth result assumes transfer optics that have not been demonstrated (Sec. IV D). At the smaller convergence angle of the myoglobin simulations, about a third of each screen leaks into the other at low spatial frequencies. Moving from phase screens at known depths to continuous specimens remains to be shown. The ZPC-level transfer of APEX needs post-specimen imaging optics again: unlike far-field 4D-STEM, APEX is subject to objective-lens aberrations at high resolution, chromatic blurring of inelastically scattered electrons, and the consequent need for zero-loss energy filtration or chromatic aberration correction[40] on thick biological specimens, just as in phase-plate CTEM. A zero-loss filter removes the inelastically scattered electrons, which 4D-STEM can record. With the inelastic mean free path that Rice *et al.*[23] measured, 395 nm, the filter removes about 2% of the incident electrons in the 6 nm of ice of our simulated specimen and 40% in 200 nm of ice. We therefore compared APEX after the filter with 4D-STEM without a filter, per incident electron (Fig. S3 (c), Table S7). In this comparison, every inelastically scattered electron counts as fully useful for 4D-STEM, which is an upper bound. Up to 50 nm of ice, APEX has the higher Fisher information in every band when both modes use the same defocus. At 100 nm, strongly defocused 4D-STEM reaches APEX in the lowest band. At 200 nm, 4D-STEM is higher in the lowest band, and the two modes are within 0.08 in the other bands. At $\alpha \approx 5$ mrad, spherical-aberration correction, in contrast, is needed only on the image side and only for high resolution; at 15 mrad, it is needed on both sides (Sec. IV C). At small $\alpha$, probe-side aberrations shift the reference and the scattered waves together, whereas image-side aberrations give the scattered waves a phase error relative to the focused spot at $C'$. At 15 mrad, the aberration phase of an uncorrected probe is about 80 times larger, and the transfer falls. The reconstruction cannot recover the image-side loss, because the Fisher information in the recorded counts is already reduced near the zeros of the image-side transfer. Russo *et al.* estimated that chromatic-aberration correction, an in-focus phase plate and liquid-helium cooling together lower the smallest protein that can be identified in situ to about 50 kDa in a 200 nm thick specimen and to about 75 kDa at 500 nm (Russo *et al.*[41]). In such volumes, APEX could add depth sectioning from a single, untilted view, which would avoid the loss of data quality with tilt that these authors emphasize. This requires that the two-screen result carries over to continuous specimens, which is not yet shown. In the two-screen model, an idealized ZPC CTEM tilt series at the same total dose separates the screens as well as APEX (Table I). The possible advantage of APEX is the untilted specimen, not a better separation. At 4 Å, the finest resolution considered by Russo *et al.*, the geometric axial half-width $0.35/(\alpha q)$ (Sec. VI) is about 26 nm at the convergence semi-angle of our myoglobin simulations. At $\alpha = 15$ mrad, the most optimistic case we simulated, it is about 9 nm. These are geometric limits for noise-free data, and they hold at this resolution only: the half-width grows in proportion to $1/q$, so that, at the $\alpha$ of Fig. 7, whole protein envelopes separate only at layer spacings of several hundred nanometres (Sec. VI B). In a 500 nm specimen, reaching these limits would also need a non-linear multislice reconstruction that accounts for the Fresnel propagation between layers, which we did not test. Our benchmark has vacuum between two phase screens, not ice or a continuous specimen. Several practical questions remain to be addressed in future work:

1. Optical design and alignment of a continuous-wave laser phase plate at the crossover image $C'$, including the required intermediate-lens magnification and positional drift tolerances.
2. Quantitative modeling of chromatic aberration, with and without chromatic-aberration correction, finite source size and partial spatial coherence.
3. Tolerance to time-dependent descan errors and drift of the element at $C'$; our estimate covers a static displacement only (Sec. IV C).
4. Iterative non-linear multislice reconstruction of thick specimens at cryo-EM doses.
5. Experimental demonstration of two-dimensional APEX at small $\alpha$ with an electrostatic element in the selected-area aperture plane of a standard microscope, and of APEX at milliradian convergence angles in a transfer-lens column (Sec. IV D).

## ACKNOWLEDGEMENTS

The author thanks Vincenzo Grillo and Mingjian Wu for discussions about the optical setup. During this work, the author used the AI coding assistant Claude Code (Anthropic) to write and test simulation and reconstruction code, to run and track computations, and to prepare figures. The author checked all code, results and text and takes full responsibility for the content of this article.

## FUNDING

This work received funding from the European Research Council (ERC) under the Horizon Europe research and innovation program (grant agreement No. 101164581, project HyperScaleEM).

## CONFLICT OF INTEREST

The author declares no conflict of interest.

## DATA AVAILABILITY STATEMENT

The simulated data, numerical results, logs and figure source data that support the findings of this study are openly available in Zenodo at doi:10.5281/zenodo.24170512. The simulation and reconstruction code is part of the open-source `scatterem` package. The scripts that run the simulations and make each figure are in the repository of this article, github.com/ECLIPSE-Lab/14_scanned_zernike. Both are released with the publication of this article.